# Transport and Magnetic properties of laser ablated $La_{0.7}Ce_{0.3}MnO_3$ films on $LaAlO_3$: Effect of oxygen pressure, sample thickness and co-doping with Ca

P. Raychaudhuri†, S. Mukherjee, A. K. Nigam, J. John, U. D. Vaisnav, R. Pinto
*Department of Condensed Matter and Materials Science, Tata Institute of Fundamental Research, Homi Bhabha Rd., Colaba, Mumbai-400005, India.*

And

P. Mandal
*Saha Institute of Nuclear Physics, Block-AF, Sector I, Salt Lake, Calcutta-700064.*

*Abstract:* $La_{0.7}Ce_{0.3}MnO_3$ is a relatively new addition in the family of colossal magnetoresistive manganites where the cerium ion is believed to be in the $Ce^{4+}$ state. In this paper we report an extensive study the magnetotransport properties of laser ablated $La_{0.7}Ce_{0.3}MnO_3$ films on $LaAlO_3$ with variation in ambient oxygen pressure during growth and film thickness. We observe that the transport and magnetic properties of the film depend on the interplay between oxygen pressure, surface morphology, film thickness and epitaxial strain. The films were characterized by x-ray diffraction on a 4-circle x-ray goniometer. We observe an increase in the metal-insulator transition temperature with decreasing oxygen pressure. This is in direct contrast with the oxygen pressure dependence of $La_{0.7}Ca_{0.3}MnO_3$ films suggesting the electron doped nature of the $La_{0.7}Ce_{0.3}MnO_3$ system. With decreasing film thickness we observe an increase in the metal-insulator transition temperature. This is associated with a compression of the unit cell in the a-b plane due to epitaxial strain. When the system is co-doped with 50% Ca at the Ce site the system ($La_{0.7}Ca_{0.15}Ce_{0.15}MnO_3$) is driven into a insulating state suggesting that the electrons generated by $Ce^{4+}$ is compensated by the holes generated by $Ca^{2+}$ valence thus making the average valence at the rare-earth site 3+ as in the parent material $LaMnO_3$.

---

† e-mail: pratap@tifr.res.in

## I. Introduction

Hole doped rare-earth manganites of the form $RMnO_3$ (R = rare-earth) have attracted a lot of attention in recent times due their interesting magnetotransport properties arising from spin charge coupling. The parent material is a charge transfer insulator[1] where the $Mn^{3+}$ moments form a layered antiferromagnetic structure. The electronic configuration of $Mn^{3+}$ is $t_{2g}^3 e_g^1$ where the three $t_{2g}$ electrons are tightly bound with a net spin of 3/2. Hole doping is achieved in these compounds by substituting a bivalent cation like Ca, Sr, Pb at the rare-earth site. Around 30% doping compounds like $La_{0.7}Ca_{0.3}MnO_3$ show ferromagnetism associated with a metal-insulator transition and large negative magnetoresistance close to the ferromagnetic transition temperature ($T_c$). These exotic properties of these material arise due to the strong on site Hund's rule coupling between the localized $t_{2g}^3$ electrons and the electrons in the $e_g$ band. This gives rise to a coupling between neighboring $Mn^{3+}/Mn^{4+}$ pairs are coupled via Zener double exchange[2] mechanism. According to this mechanism the hopping probability of an electron between two neighboring $Mn^{3+}/Mn^{4+}$ pairs is proportional to $\cos(\theta/2)$[3]. $La_{0.7}Ce_{0.3}MnO_3$ is unique in this family of compounds since the trivalent La is replaced by Ce instead of a bivalent cation. This compound has orthogonal structure with $a \approx b \approx \sqrt{2}c$. This compound shows metal-insulator transition[4,5] and ferromagnetism associated with large negative magnetoresistance similar to $La_{0.7}Ca_{0.3}MnO_3$. To explain the observed behavior P. Mandal and S. Das[6] have argued that Ce could exist in in $Ce^{4+}$ state like in $CeO_2$ thus doping electrons instead of holes in the parent material. Thus according to their picture the double exchange in this compound operates between $Mn^{2+}/Mn^{3+}$ neighboring pairs giving rise to metal insulator transition and ferromagnetism. From the theoretical point of view the double exchange mechanism is symmetric with respect to electron or hole doping in the parent compound. Incidentally, some of the present authors had reported metal-insulator transition and ferromagnetism in the electron doped layered manganite $La_{1.8}Y_{0.5}Ca_{0.7}Mn_2O_7$ [7,8].

In this paper we report a detail study of $La_{0.7}Ce_{0.3}MnO_3$ films prepared by pulsed laser ablation on $LaAlO_3$. In some cases we also prepared $La_{0.7}Ca_{0.3}MnO_3$ films for comparison with these films. We compare the effect of ambient oxygen pressure during growth on $La_{0.7}Ce_{0.3}MnO_3$ and $La_{0.7}Ca_{0.3}MnO_3$ films. To address the question of the valence state of Ce we prepared films of nominal composition $La_{0.7}(Ce_{0.15}Ca_{0.15})MnO_3$. This compound goes into an insulating state when the film is grown at 100mTorr oxygen pressure suggesting that the electrons generated by $Ce^{4+}$ compensate the holes generated by $Ca^{2+}$. Finally, we study the effect epitaxial strain on the metal-insulator transition temperature ($T_p$) by varying the film thickness.

## II. Experiment

Films of $La_{0.7}Ce_{0.3}MnO_3$, $La_{0.7}Ca_{0.3}MnO_3$ and $La_{0.7}Ce_{0.15}Ca_{0.15}MnO_3$ were prepared by pulsed laser deposition using a KrF eximer laser in oxygen atmosphere. The substrate temperature was kept between 750-770$^0$C for all the films. The laser energy density was kept at roughly 160mJ/mm$^2$ per shot with a repetition rate of 10Hz. Films were grown at 3 oxygen pressures: 100mTorr, 200mTorr, 400mTorr. Films used for thickness dependence study were all grown at 400mTorr. After deposition the laser ablation chamber was vented with high purity oxygen and the substrate cooled down to room temperature. The lattice parameters of the films by x-ray diffraction on a high resolution 4-circle x-ray goniometer.

Film thickness was measured using stylus method on a Dektak profilometer. Resistance and magnetoresistance were measured by conventional 4-probe technique. Magnetization was measured on a Quantum Design superconducting quantum interference device (SQUID) magnetometer. Surface morphology was checked using a Digital Instrument atomic force microscope.

## III. Results and discussion

**A. Effect of ambient oxygen pressure during growth:**

To study the effect of oxygen pressure on metal insulator transition ($T_c$) and the ferromagnetic transition ($T_p$) films of $La_{0.7}Ce_{0.3}MnO_3$ were grown at 3 different oxygen pressures: 100mTorr, 200mTorr and 400mTorr. Other parameters like deposition time, laser energy, and substrate temperature were kept constant for all the films. For comparison the $La_{0.7}Ca_{0.3}MnO_3$ films were also grown under the same conditions. All the films were seen to have c⊥(film plane) from the x-ray $\theta-2\theta$ scan. Figure 1 shows a representative x-ray diffraction $\theta-2\theta$ scans of the $La_{0.7}Ce_{0.3}MnO_3$ film grown at 400mTorr showing the (00n) peaks. Figures 2a-f show the atomic force microscope (AFM) image of the three $La_{0.7}Ce_{0.3}MnO_3$ and $La_{0.7}Ca_{0.3}MnO_3$ film surfaces. There is a degradation of the surface morphology and grain size for the $La_{0.7}Ce_{0.3}MnO_3$ film grown at low oxygen pressure (100mTorr) (figure 2c). This degradation in the surface morphology at low oxygen pressure is also observed in $La_{0.7}Ca_{0.3}MnO_3$ films grown at 100mTorr (figure 2f). Figures 3a-c show the resistance versus temperature (R-T) from for the three films. The interesting fact is that the metal-insulator transition temperature ($T_p$) and the ferromagnetic transition temperature ($T_c$) increases from 246K for the film grown at 400mTorr to 272K for the film grown at 100mTorr. However, the effect of ambient oxygen pressure on $T_p$ in $La_{0.7}Ce_{0.3}MnO_3$ is in direct contrast to the effect of ambient oxygen pressure on $T_p$ in $La_{0.7}Ca_{0.3}MnO_3$. Figures 3d-f show the R-T curves for $La_{0.7}Ca_{0.3}MnO_3$ films grown at the same pressures. In the case of $La_{0.7}Ca_{0.3}MnO_3$ the metal-insulator transition broadens drastically with decreasing oxygen pressure with a small decreasing trend in $T_p$. Before elaborating further on these results we shall discuss the effect of co-doping Ce and Ca in $La_{0.7}Ce_{0.15}Ca_{0.15}MnO_3$. However, we would like to point out that this property might be useful from the technological point of view since this allows $La_{0.7}Ce_{0.3}MnO_3$ films to be synthesized on a wider range of oxygen pressures without compromising on the metal-insulator transition width.

In spite of the initial picture proposed in ref.6 electron doping in $La_{0.7}Ce_{0.3}MnO_3$ via $Ce^{4+}$ valence state has been a questionable issue since the data of thermo-electric power (TEP) in this material (bulk) is similar to that of hole doped manganites. However, the analysis TEP data in these materials could be very complex, in particular where the TEP changes sign as a function of temperature[6]. To resolve this problem we prepared films of the nominal composition $La_{0.7}Ce_{0.15}Ca_{0.15}MnO_3$. If Ce is in the 4+ state one would expect the holes generated by $Ca^{2+}$ be compensated by the electrons generated by $Ce^{4+}$ thus driving the system into an insulating state like in the parent compound $LaMnO_3$. The difficulty in this is that the undoped compound (like $LaMnO_3$) has a tendency to get overoxygenated, generating cation vacancies[9,10,11]. It has been shown that the overoxygenated compound can show a metal-insulator transition just like the doped material[12]. However, if the compound $La_{0.7}Ce_{0.15}Ca_{0.15}MnO_3$ is compensated for electrons and holes the effect of varying ambient oxygen pressure during growth should be much more drastic than in $La_{0.7}Ca_{0.3}MnO_3$ or $La_{0.7}Ce_{0.3}MnO_3$ films since one should be able to suppress the carries by decreasing the oxygen pressure. In figure 4a-b we show the R-T curve for $La_{0.7}Ce_{0.15}Ca_{0.15}MnO_3$ films grown at 400mTorr and 100mTorr oxygen pressure. The film grown at 400mTorr shows a metal-insulator transition around 230K. However, the film grown at 100mTorr becomes an insulator with resistance 4 orders of magnitude larger than the films grown at 400mTorr at 55K. Comparing this result with the effect of varying oxygen pressure in the same range in $La_{0.7}Ca_{0.3}MnO_3$ and $La_{0.7}Ce_{0.3}MnO_3$ (figure 3) we see that the low temperature metallic phase in these two compounds is never fully suppressed. The drastic increase in the resistance and the suppression of metallicity in the $La_{0.7}Ce_{0.15}Ca_{0.15}MnO_3$ film grown at 100mTorr thus indicates that electrons and holes are compensated in this compound.

Having thus established that Ce is in 4+ state one can now look more carefully at the data on $La_{0.7}Ce_{0.3}MnO_3$ and $La_{0.7}Ca_{0.3}MnO_3$ in figure 3. Firstly the increase in $T_p$ with decreasing oxygen pressure in $La_{0.7}Ce_{0.3}MnO_3$ could be understood as follows. The parent compound $LaMnO_3$ the Mn ion has a filled $t_{2g}$ orbital with three electrons and one electron at the $e_g$ orbital. The average occupancy at the $e_g$ orbital (n) is therefore 1. The twofold degeneracy of the $e_g$ orbital is lifted in due to the Jahn-Teller distortion in $LaMnO_3$ making the material an insulator. In $La_{0.7}Ca_{0.3}MnO_3$ the parent compound is doped with holes and the average occupancy $0<n<1$. On the other hand in $La_{0.7}Ce_{0.3}MnO_3$ the parent compound is doped with electrons and $1<n<2$. Since the oxygen content of the film depends on the ambient oxygen pressure during growth we assume the $La_{0.7}Ce_{0.3}MnO_3$ film to have the composition $La_{0.7}Ce_{0.3}MnO_{3-\delta}$. (We wish to point out that at this point we do not attach too much significance to the negative sign of $\delta$ since it is not known whether this material gets overoxygenated or under oxygenated. $\delta$ is used at this point to relatively compare films with different oxygen stoichiometry). For this composition the average valence of Mn is $2.7-2\delta$. This means that when $\delta$ is increased the average manganese valence reduces towards 2. This in turn increases the electron doping at the $e_g$ orbital. This is in contrast to the situation in the hole doped $La_{0.7}Ca_{0.3}MnO_{3-\delta}$. In this case the average Mn valence is $3.3-2\delta$. Here increasing $\delta$ means that the Mn valence will reduce towards 3 thus decreasing hole doping. Thus in terms of doping the reducing oxygen pressure has opposite effect in the two compounds. This is schematically shown in figure 5. (The $e_g$ band in these compounds are spin split due to the large on-site Hund's rule coupling.) Since $T_p$ and $T_c$ are known to be sensitive function of doping in these materials this could give rise to increase in $T_p$ in one case and decrease in the other. The other difference between $La_{0.7}Ca_{0.3}MnO_3$ and $La_{0.7}Ce_{0.3}MnO_3$ films is that the width of the metal-insulator transition is much less affected by the decrease in the oxygen pressure in $La_{0.7}Ce_{0.3}MnO_3$. For the $La_{0.7}Ca_{0.3}MnO_3$ film grown at 100mTorr the metal-insulator transition becomes extremely wide (figure 3f). On the other hand the width of the transition remains almost constant (with a slight decrease at lower oxygen pressures) with decrease in oxygen pressure for $La_{0.7}Ce_{0.3}MnO_3$ film (figure 3g). The large broadening in the metal-insulator transition in $La_{0.7}Ca_{0.3}MnO_3$ suggest that the $La_{0.7}Ca_{0.3}MnO_3$ films becomes inhomogeneously oxidized at lower oxygen pressures. We believe that the $La_{0.7}Ce_{0.3}MnO_3$ film is more homogeneous in terms of its oxygen stoichiometry due to the high oxygen affinity of cerium. This robustness of the metal-insulator transition might be advantageous from the application point of view where a sharp metal-insulator transition and a large temperature coefficient of resistivity is desirable, for example, in the case of bolometric application of perovskite manganite films[13,14].

**B. The effect of epitaxial strain on the metal-insulator transition ($T_p$) and magnetoresistance:**

In order to study the effect of epitaxial strain on the transport properties of this material we synthesized films in of varying thickness in the range 500-3200 Å. For the thinnest film the thickness was estimated from the deposition time. The substrate temperature was kept at $760^0$C and the ambient oxygen pressure was kept at 400mTorr during growth. These values were chosen since grain size and surface morphology was best for these parameters. Figure 6a-c show the surface morphology for films of various thicknesses. Thicker films have smaller grain size possibly due to strain relaxation. In figure 7a we show the R-T curves close to the metal insulator transition. There is a gradual increase in $T_p$ with decreasing film thickness. However we see that there is not much change in the width of the metal-insulator transition in this thickness range. This is shown in figure 7b where $R/R_{max}$ is plotted as a function of $T/T_p$. Figure 7c shows the magnetoresistance (MR~$\Delta\rho/\rho_0$~$(\rho(H)-\rho(0))/\rho(0)$) at 300 K for the four samples. There is a gradual increase in the magnetoresistance with decreasing film thickness.

In order to understand this phenomenon we measured the lattice constant of the 4 films using X-ray diffraction on a 4-circle high resolution X-ray goniometer. The lattice constants along with the cell volume a $T_p$ for films of various thickness are summarized in Table I. There is a gradual compression in the in-plane (a-b) lattice constants with decreasing thickness associated with a expansion in the c-axis lattice parameter. This is understandable since the pseudo-cubic lattice parameter of $LaAlO_3$ is smaller than the pseudo-cubic lattice parameters of $La_{0.7}Ce_{0.3}MnO_3$. To understand the increase in $T_p$ with decrease in thickness we note that $T_p$ in the bulk compound increases with applied pressure[15]. It has been shown that $T_p$ (and $T_c$) in doped manganites is governed by by the bandwidth (W) of the $e_g$ band and is given by the semi empirical formula[16]

$$W \propto \frac{\cos\omega}{d_{Mn-O}^{3.5}},$$

where $\omega$ is the "tilt" angle in the plane of the Mn-O-Mn bond (given by $\omega = \frac{1}{2}(\pi - \langle Mn\text{-}O\text{-}Mn \rangle)$), $d_{Mn-O}$ is the Mn-O bond length. The applied isotropic pressure in bulk samples reduces $d_{Mn-O}$ but keeps $\omega$ almost constant thus increasing W and $T_c$. On the other hand when a chemical pressure is applied by modifying the size of the cation at the rare-earth site $\cos\omega$ decreases more rapidly than $d_{Mn-O}$ thus reducing W and $T_c$. We believe that the effect of compressive epitaxial stress in films is similar to the applied pressure in the bulk sample. However the situation is more complicated here due to anisotropic in-plane and out of plane strain in the film which give rise to different values of $d_{Mn-O}$ and $\omega$ in different directions. R. A. Rao et al[17] have reported the effect of epitaxial strain in $La_{0.8}Ca_{0.2}MnO_3$ films of varying thickness. Their study revealed that $T_p$ decreases with compressive in-plane epitaxial strain. The puzzling fact is that in bulk $La_{0.7}Ca_{0.3}MnO_3$ also $T_p$ increases with applied pressure[16]. M. G. Blamire et al[18] have reported similar results on laser ablated $La_{0.7}Ca_{0.3}MnO_3$ films on $LaAlO_3$ and $SrTiO_3$. A close inspection of their R-T data for various film thicknesses reveal that the metal-insulator transition becomes extremely broad in addition to the decrease in $T_p$ with decreasing film thickness. One reason for this behavior could be that very thin films of $La_{0.7}Ca_{0.3}MnO_3$ tend to lose oxygen thereby reducing its $T_p$[17].

Figure 7a-d show the R-T curves close to the metal-insulator transition for the films with varying thickness at different fields up to 1.5T. All the films have a large MR > 60% close to the metal-insulator transition at 1.5T (Figure 7e). One point to note here is that above 250K thinner films have larger MR. This could be useful from the application point of view since a large MR at temperatures close to room temperature is required for the device application of these films.

## IV. Conclusion

In summary, we have reported in this paper the magnetic and transport properties of laser ablated $La_{0.7}Ce_{0.3}MnO_3$ films on $LaAlO_3$ substrates. By co-doping with Ca in $La_{0.7}Ca_{0.15}Ce_{0.15}MnO_3$ we have established that in this system the parent compound ($LaMnO_3$) is doped with electrons in contrast to the hole doped compounds like $La_{0.7}Ca_{0.3}MnO_3$ where holes are doped by substitution of bivalent $Ca^{2+}$ at the rare-earth site. We have compared the effect of ambient oxygen pressure during growth in $La_{0.7}Ca_{0.3}MnO_3$ and $La_{0.7}Ce_{0.3}MnO_3$. We find that the latter is much more stable with respect to variation in oxygen pressure. Thickness dependence studies reveal that the metal-insulator transition temperature increases with decreasing thickness. This can be attributed to the compressive in-plane epitaxial strain in film. In conclusion, the properties of $La_{0.7}Ce_{0.3}MnO_3$ provide a scheme to tune the metal-insulator transition temperature over a moderate range in this colossal magnetoresistive manganite by varying oxygen pressure during growth and sample thickness without compromising on the sharpness of the metal-insulator transition.

## **Acknowledgement**

We would like to thank G.S.Nathan for being involved in a part of this work. We also thank R.S.Sannabhadti and Arun Patade for technical support.

**Figure Captions:**

Figure 1: X-ray $\theta-2\theta$ scan of the $La_{0.7}Ce_{0.3}MnO_3$ film grown at 400mTorr showing the (00n) peaks of the orthorhombic structure.

Figure 2: AFM photograph of the $La_{0.7}Ce_{0.3}MnO_3$ films grown at (a) 400mTorr, (b) 200mTorr, (c) 100mTorr. AFM photograph of the $La_{0.7}Ca_{0.3}MnO_3$ films grown at (d) 400mTorr, (e) 200mTorr, (f) 100mTorr.

Figure 3: Resistance versus temperature of $La_{0.7}Ce_{0.3}MnO_3$ grown at (a) 400 mTorr, (b) 200mTorr, (c) 100mTorr. Resistance versus temperature of $La_{0.7}Ca_{0.3}MnO_3$ grown at (d) 400 mTorr, (e) 200mTorr, (f) 100mTorr; (g) shows the $R/R_{max}$ versus $T/T_p$ for the three $La_{0.7}Ce_{0.3}MnO_3$ films where $R_{max}$ is the resistance value at $T_p$.

Figure 4: Resistance versus temperature for $La_{0.7}Ca_{0.15}Ce_{0.15}MnO_3$ films grown at (a) 400 mTorr and (b) 100mTorr oxygen pressure.

Figure 5: Schematic band structure of $La_{0.7}Ca_{0.3}MnO_{3-\delta}$ and $La_{0.7}Ce_{0.3}MnO_{3-\delta}$. The Ferni energy $E_F$ increases with in both cases with increase in $\delta$. However this corresponds to decrease in hole doping in $La_{0.7}Ca_{0.3}MnO_{3-\delta}$ whereas in $La_{0.7}Ce_{0.3}MnO_{3-\delta}$ this increases the electron dpoing in the parent material.

Figure 6: Surface morphology of $La_{0.7}Ce_{0.3}MnO_3$ films of various thickness: (a) 500 Å (b) 1000 Å (c) 3200 Å.

Figure 7: (a) Resistance versus temperature around $T_p$ for films of various thickness; (b) $R/R_{max}$ versus $T/T_p$ for films of various thickness; (c) MR versus field at 300 K for films of various thickness.

Figure 8: (a)-(d) Resistance versus temperature at different fields for various film thickness. (e) MR at 1.5 T as a function of temperature for films of various thickness.

**Table Caption**

Table I: Evolution of lattice parameters, cell volume and $T_p$ with thickness variation in $La_{0.7}Ce_{0.3}MnO_3$ films.

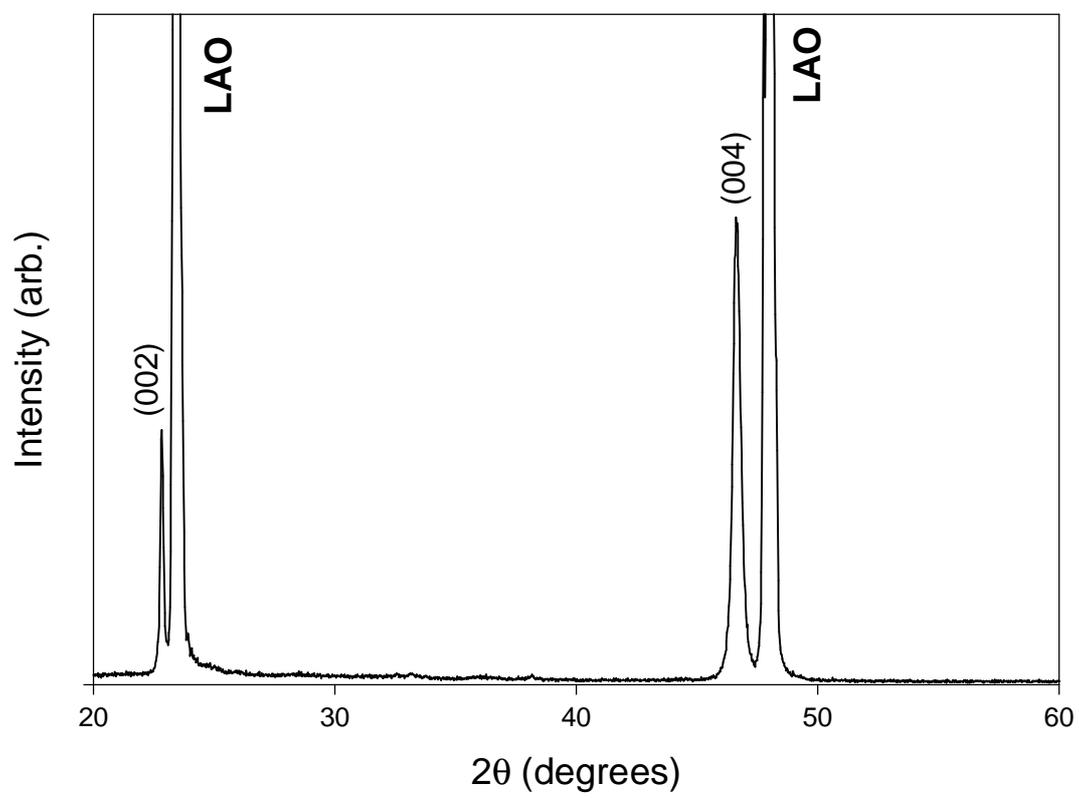

Figure 1 (P Raychaudhuri et.al.)

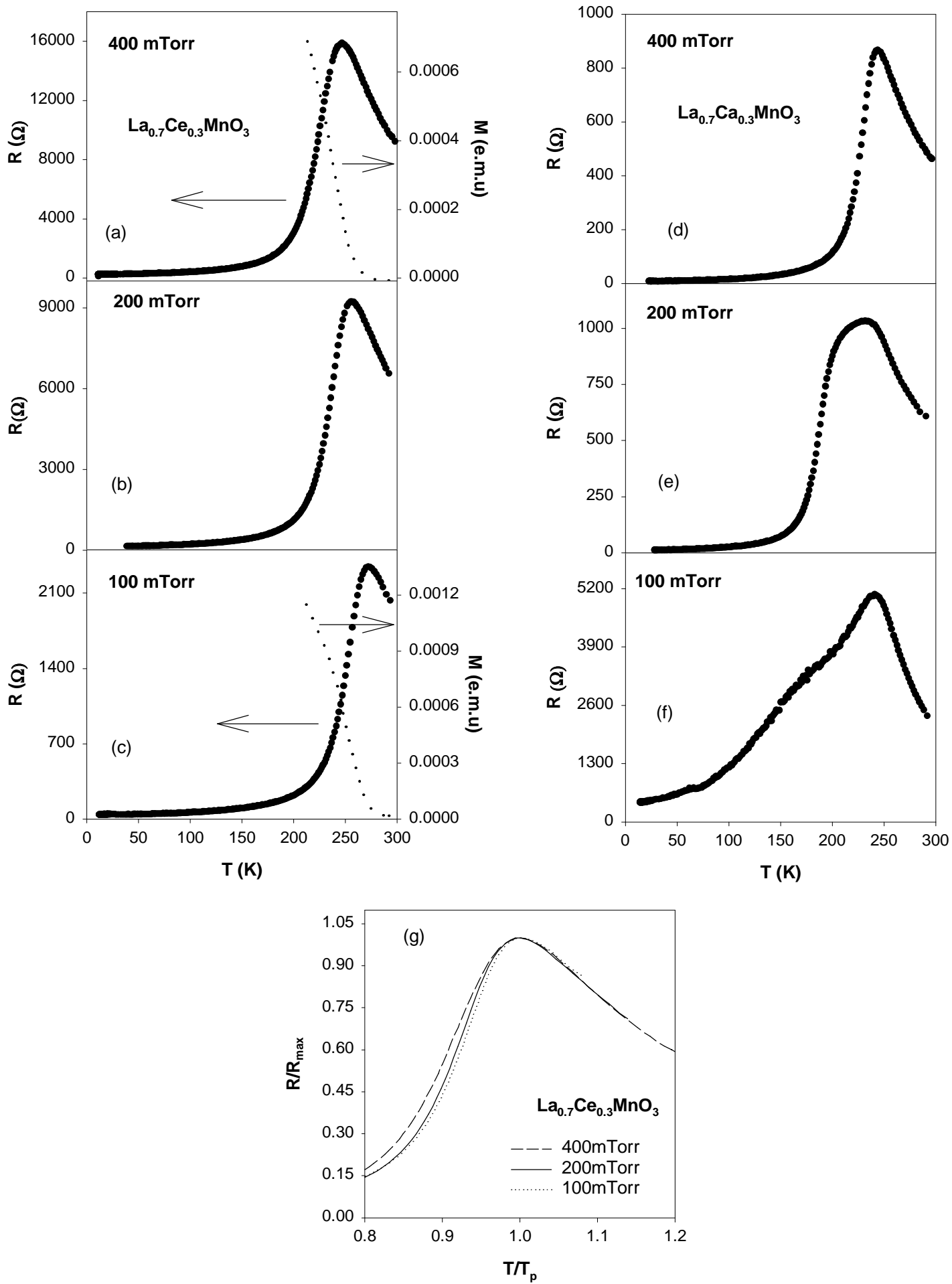

Figure 3 (P. Raychaudhuri et al)

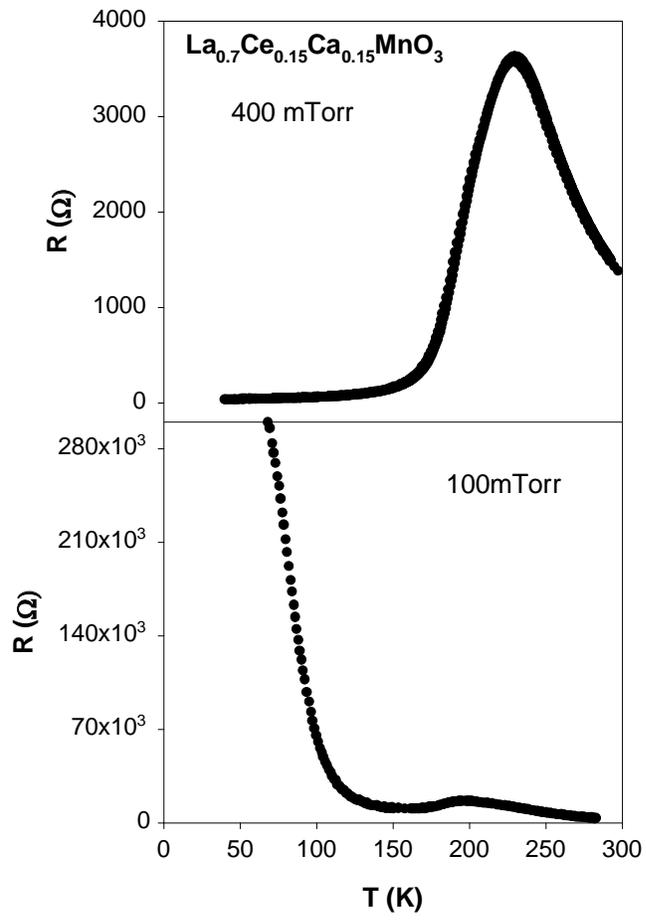

Figure 4 (P Raychaudhuri et al)

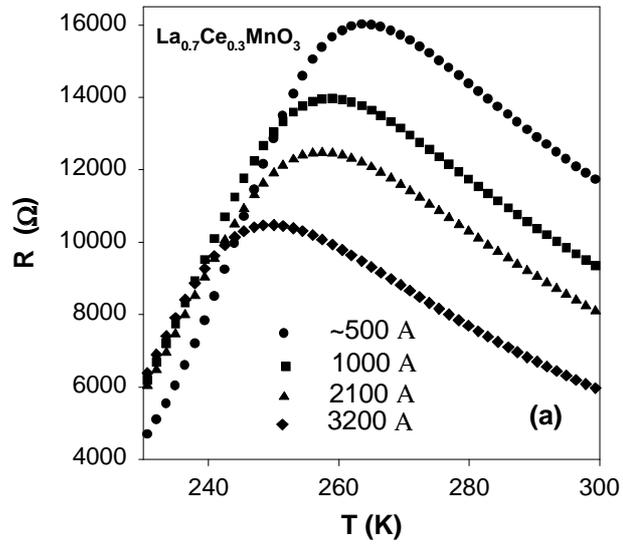

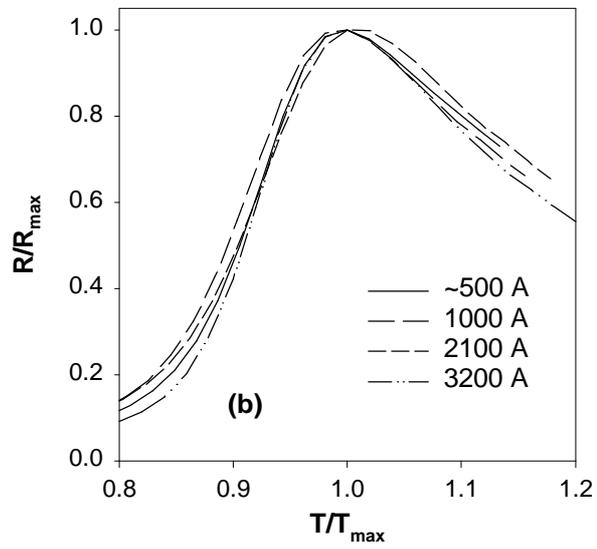

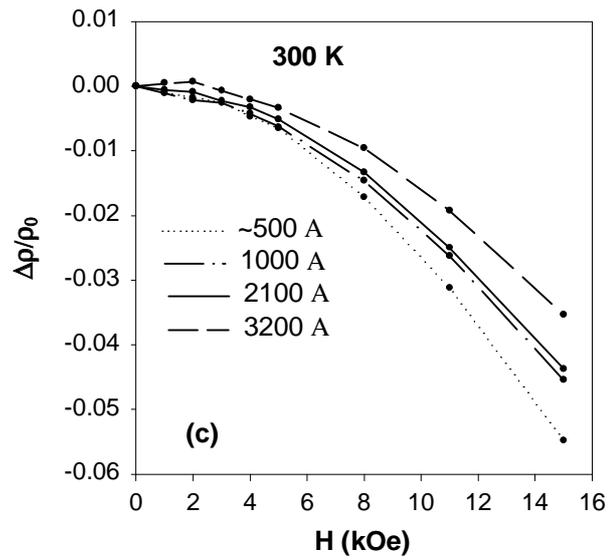

**Figure 7 (P Raychaudhuri et al)**

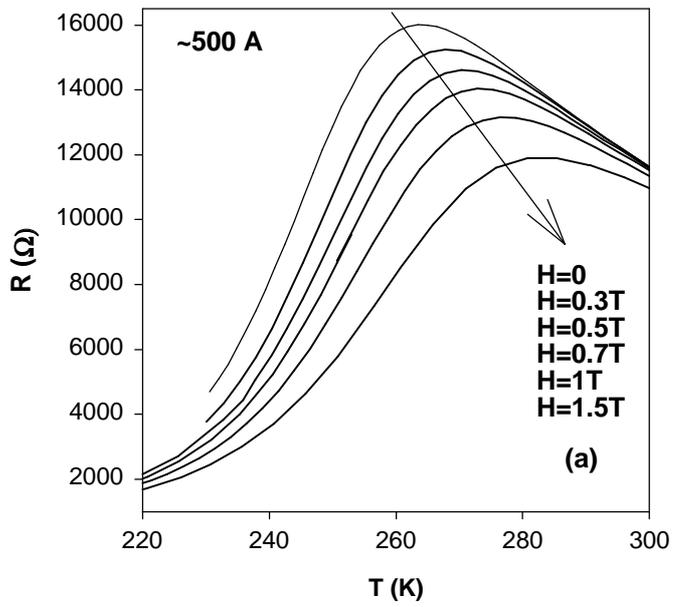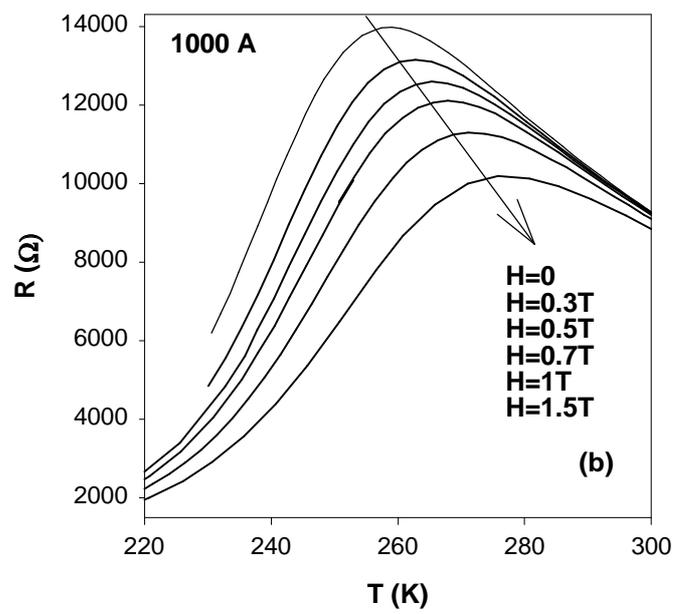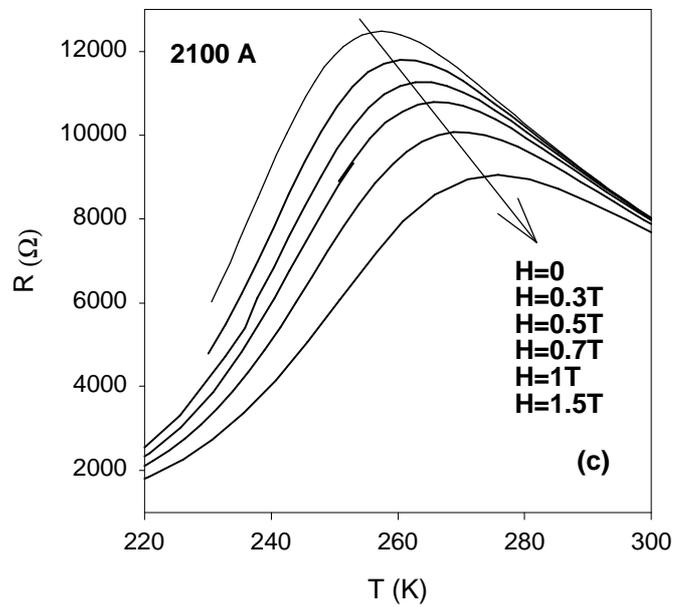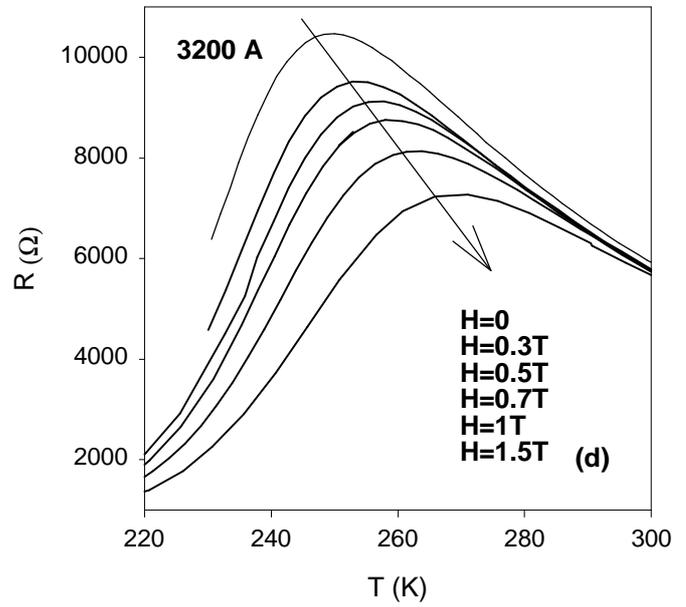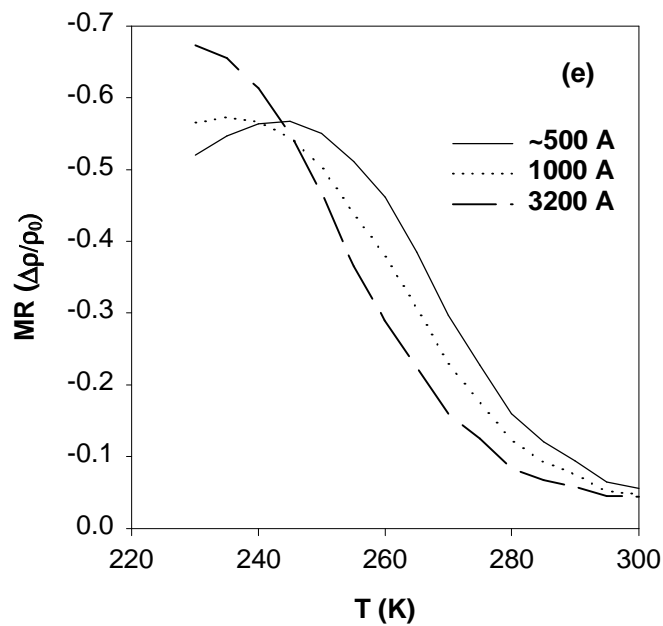

Figure 8 (P Raychaudhuri et al)

**Table I**

| Thickness[*] (Å) | a (Å) | b (Å) | c (Å) | V (Å³) | $T_p$ (K) |
|---|---|---|---|---|---|
| 500 | 5.518 | 5.459 | 7.852 | 236.52 | 263 |
| 1000±100 | 5.529 | 5.466 | 7.841 | 236.09 | 259 |
| 2100±100 | 5.548 | 5.464 | 7.836 | 237.54 | 257 |
| 3200±100 | 5.561 | 5.512 | 7.802 | 239.15 | 250 |

[*]All thickness values except the lowest one were measured on a Dektak profilometer. The lowest thickness was estimated from the time of deposition using the calibration from the thicker ones.

P Raychaudhuri et al